\documentclass[%
 aip,
 amsmath,amssymb,
 reprint,%
,jap]{revtex4-1}

\usepackage{graphicx}
\usepackage{dcolumn}
\usepackage{bm}

\usepackage[utf8]{inputenc}
\usepackage[T1]{fontenc}
\usepackage{mathptmx}

\begin{document}

\preprint{AIP/123-QED}

\title[Optomechanical resonance]{Out-of-equilibrium optomechanical resonance self-excitation}

\author{P. Milde}
\affiliation{Institut f\"ur Angewandte Physik, Technische Universit\"at Dresden, Noethnitzer-Strasse 61, 01187 Dresden, Germany
}%

\author{M. Langenhorst}%
\affiliation{ 
Physikalisches Institut, Karlsruher Institut f\"ur Technologie, D-76128 Karlsruhe, Germany}%

\author{H. H\"olscher}
\affiliation{%
Institute for Microstructure Technology (IMT), Karlsruhe Institute of Technology, Hermann-von-Helmholtz-Platz 1, 76344 Eggenstein-Leopoldshafen, Germany}%

\author{J. Rottmann-Matthes}
\affiliation{%
Institute for Analysis, Karlsruhe Institute of Technology, Englerstrasse 2, 76131 Karlsruhe, Germany}%

\author{D. Hundertmark}
\affiliation{%
Institute for Analysis, Karlsruhe Institute of Technology, Englerstrasse 2, 76131 Karlsruhe, Germany}%

\author{L. M. Eng}
\affiliation{Institut f\"ur Angewandte Physik, Technische Universit\"at Dresden, Noethnitzer-Strasse 61, 01187 Dresden, Germany
}%

\author{R. Hoffmann-Vogel}
 \email{hoffmannvogel@uni-potsdam.de}
\affiliation{%
Department of Physics, University of Potsdam, Karl-Liebknecht Stra{\ss}e 24-25,
14476 Potsdam-Golm, Germany}%

\date{\today}

\begin{abstract}
The fundamental sensitivity limit of atomic force microscopy is strongly correlated to the thermal noise of the cantilever oscillation. A method to suppress this unwanted noise is to reduce the bandwidth of the measurement, but this approach is limited by the speed of the measurement and the width of the cantilever resonance, commonly defined through the quality factor $Q$. However, it has been shown that optomechanical resonances in interferometers might affect the cantilever oscillations resulting in an effective quality factor $Q_{\rm eff}$. When the laser power is sufficiently increased the cantilever oscillations might even reach the regime of self-oscillation. In this self-oscillation state, the noise of the system is partially determined by the interaction with the laser light far from equilibrium. Here, we show and discuss how tuning of the laser power leads to nonlinear optomechanical effects that can dramatically increase the effective quality factor of the cantilever leading to out-of-equilibrium noise. We model the effects using a fourth order nonlinearity of the damping coefficient.
\end{abstract}

\maketitle

\section{\label{sec:level1}Introduction}

In atomic force microscopy (AFM) \cite{binnig86p1} thermal noise of the cantilever constitutes the fundamental detection limit. Atomic resolution was achieved more than two decades ago in vacuum by utilising dynamic modes where the cantilever is oscillated close to the sample surface \cite{giessibl95p1, kitamura95p1}. This breakthrough became possible through the application of the so-called frequency-modulation mode \cite{albrecht91p1} where the cantilever was originally self-oscillated. Since the cantilever is not damped by a surrounding medium, the $Q$-factor of the cantilever in vacuum is large ($Q \propto 10 000$) compared to air ($Q \propto 100$) or liquids ($Q \propto 1$), because in vacuum the cantilever is damped by internal damping only. In this case, the frequency-modulation detection scheme has some advantages compared to the amplitude-modulation detection scheme and the thermal noise of the frequency is given by \cite{albrecht91p1} 

\begin{equation}
    \Delta f_{\mbox{\scriptsize rms}}=\frac{1}{A}\sqrt{\frac{k_B T B f_0}{\pi\, Q\, c_L}},
    \label{eqn-noise}
\end{equation}

\noindent
where $\Delta f_{\rm rms}$ is the root mean square value of the frequency, $A$ the oscillation amplitude, $B$ the detection bandwidth, and $c_{\rm L}$ the longitudinal spring constant of the cantilever. As the quality factor strongly influences this fundamental noise limit it seems highly advantageous to increase the $Q$-factor of the cantilever and several options to achieve an increase have been published  \cite{gysin04p1,lubbe10p1}.

\begin{figure}
\includegraphics[width=\linewidth]{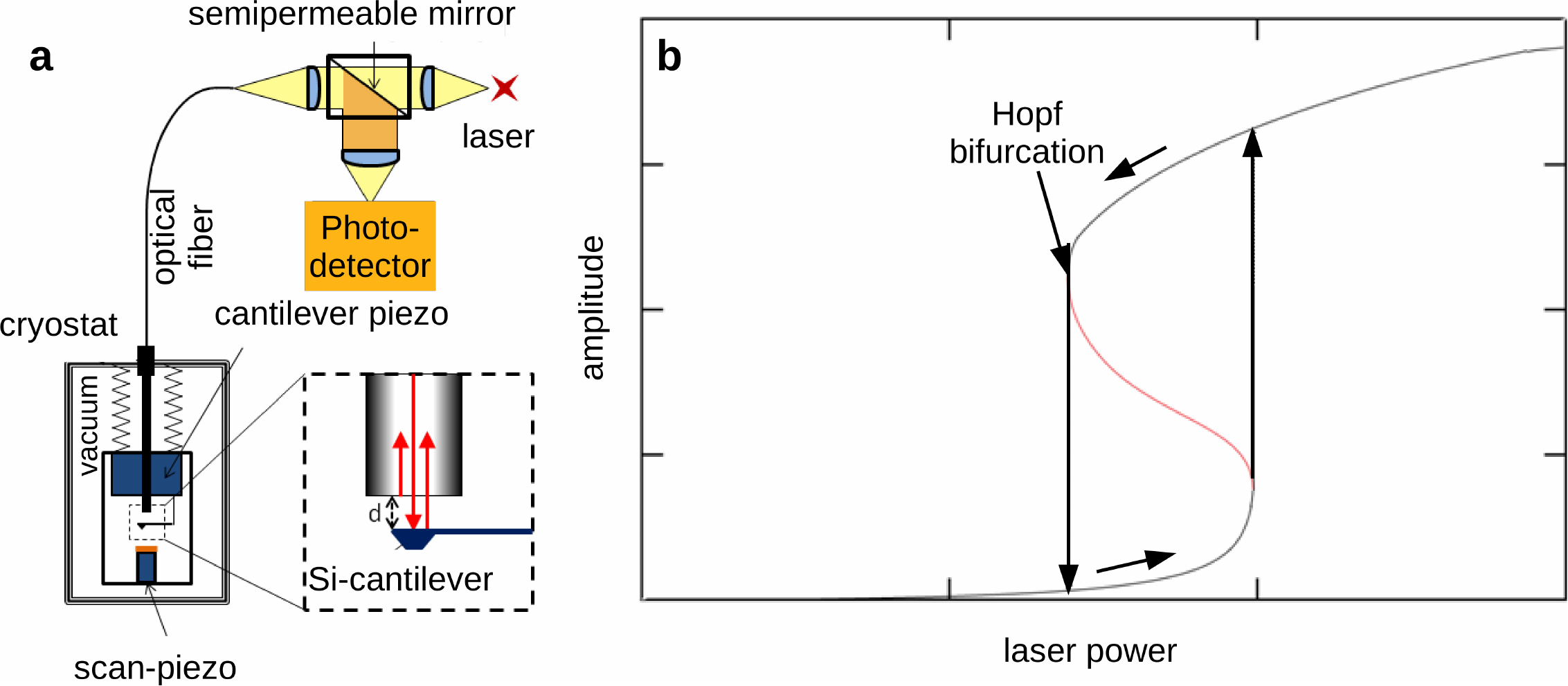}
\caption{\label{fig:SetUp} a) Schematic of the atomic force microscope set-up with interferometer detection used in this study (not to scale). 
b) Schematic of the amplitude hysteresis observed during the in- and decrease of the laser power due to bistable states. During the decrease of the laser power, near the point where the amplitude decreases again and falls back to the low-amplitude state a critical point related to a Hopf bifurcation is expected.}
\end{figure}

Soon after the invention of the AFM in 1986 it become clear that interferometer detection (Fig. \ref{fig:SetUp}a) is the method of choice to achieve highest accuracy to record the cantilever position, in particular under ultra-high vacuum (UHV), low temperature conditions \cite{martin87p1, schoenenberger89p1, rugar89p1, moser93p1}. In most set-ups the fiber-cantilever system constitutes a Fabry-P\'erot cavity usually of a rather low finesse. In such a system, photothermal forces or radiation pressure lead to a dynamic back-action \cite{kippenberg08p1, gigan06p1,pan18p1}. Since the mass of the tip and the cantilever is small, mutual influences of the light wave and the cantilever oscillations are possible \cite{schmidsfeld15p1} causing optomechanical resonances \cite{dorsel83p1, rao91p1, aspelmeyer14p1, metcalfe14p1}. Changes of the cantilever’s effective quality factor depending on its position with respect to the interferometer fringes were reported \cite{hoelscher09p1} and it has also been predicted \cite{zaitsev12p1,girvin06p1} and observed \cite{dorsel83p1,fu11p1} that optomechanical resonances lead to self-excitation and hysteresis of the noise spectrum as a function of laser power, for a recent overview see \cite{miller18p1}. As the laser power is increased, the effective quality factor of the cantilever reaches a critical point where $1/Q_{\rm eff}$ first becomes zero and then negative, leading to a non-steady state where an oscillation builds up. The amplitude of the oscillation is limited by non-linear effects resulting in our case from a force exponentially increasing with time delay. A new steady state with positive effective quality factor and non-zero oscillation amplitude is reached. As the laser power is decreased, the oscillation amplitude remains large, and the effective quality factor again reaches a critical point where $1/Q_{\rm eff}$ becomes near zero, i.e. $Q_{\rm eff}$ shows a maximum. Shortly below this point, the oscillation amplitude nearly vanishes and the effective quality factor becomes positive again. Since the first critical laser power is larger than the second one, hysteresis is observed.

Mathematically, the hysteresis is governed by two stable and one unstable solution related to a subcritical Hopf bifurcation \cite{zaitsev12p1}. Radiation pressure and photothermal forces introduce both, nonlinear forces and damping terms of any order. Previous models \cite{zaitsev12p1,girvin06p1} have focussed on nonlinear forces with constant damping rather than on nonlinear damping. At large oscillation amplitude a more than linear increase of the damping, or nonlinear forces, both stop the oscillation amplitude from increasing further. The schematic in Fig.\,\ref{fig:SetUp}b) shows the hysteresis loop with jumps of the oscillation amplitude occurring when the laser power density is increased or lowered beyond a critical value. The effects of dynamic back-action are sometimes referred to as laser "cooling", or, in the case discussed here, laser "heating" \cite{poot12p1}. Laser "cooling" is understood as a reduction of the amplitude noise, i.e. fluctuations, due to the interaction with the laser. The noise spectrum then becomes flatter and resembles one of a lower temperature. Upon "heating", amplitude fluctuations are amplified up the point where the system self-oscillates. In this heated state, the noise of the system is out of equilibrium in a peculiar way. In the simplest case, cooling is observed for a negative slope of the interferometer while heating is observed for a positive slope \cite{hoelscher09p1}. The situation is more complicated if Lissajous figures are observed \cite{flaeschner15p1}. Laser cooling has been used to cool a mechanical resonator into its quantum-mechanical ground state \cite{chan11p1}. Laser cooling methods have been developed where the incident laser light is tuned to the motional sidebands of the optical resonance of the cavity \cite{schliesser08p1}. For these methods, a large quality factor of both optical and mechanical modes is important. So far, mainly radiation pressure forces have been investigated with strong nonlinear effects. Mainly high-finesse optomechanical cavities have been used for measurements.

Here, we consider the case of an AFM in the state where it can be used for AFM measurements. The finesse of its cavity is low, and the photothermal force dominates over radiation pressure. We show that even in this situation, nonlinear effects are important enough to cause hysteresis. We demonstrate that the effective $Q$-factor of the cantilever dramatically increases as the laser power is lowered on that branch of the hysteresis loop. We observe that near the critical point related to the Hopf bifurcation, the effective $Q$-factor is strongly increased and the thermally excited spectrum of the cantilever motion reflects an out-of-thermal-equilibrium oscillation state of the cantilever. The general behavior of the system can be described mathematically by using a fourth order dependence of the damping coefficient on displacement. This allows us to consider the influence of the nonlinear effects on damping directly in the model.

\section{\label{sec:level2}Experimental methods}

\subsection{\label{subsec:21}Interferometer and cantilever}

\begin{figure}
\includegraphics[width=0.9\linewidth]{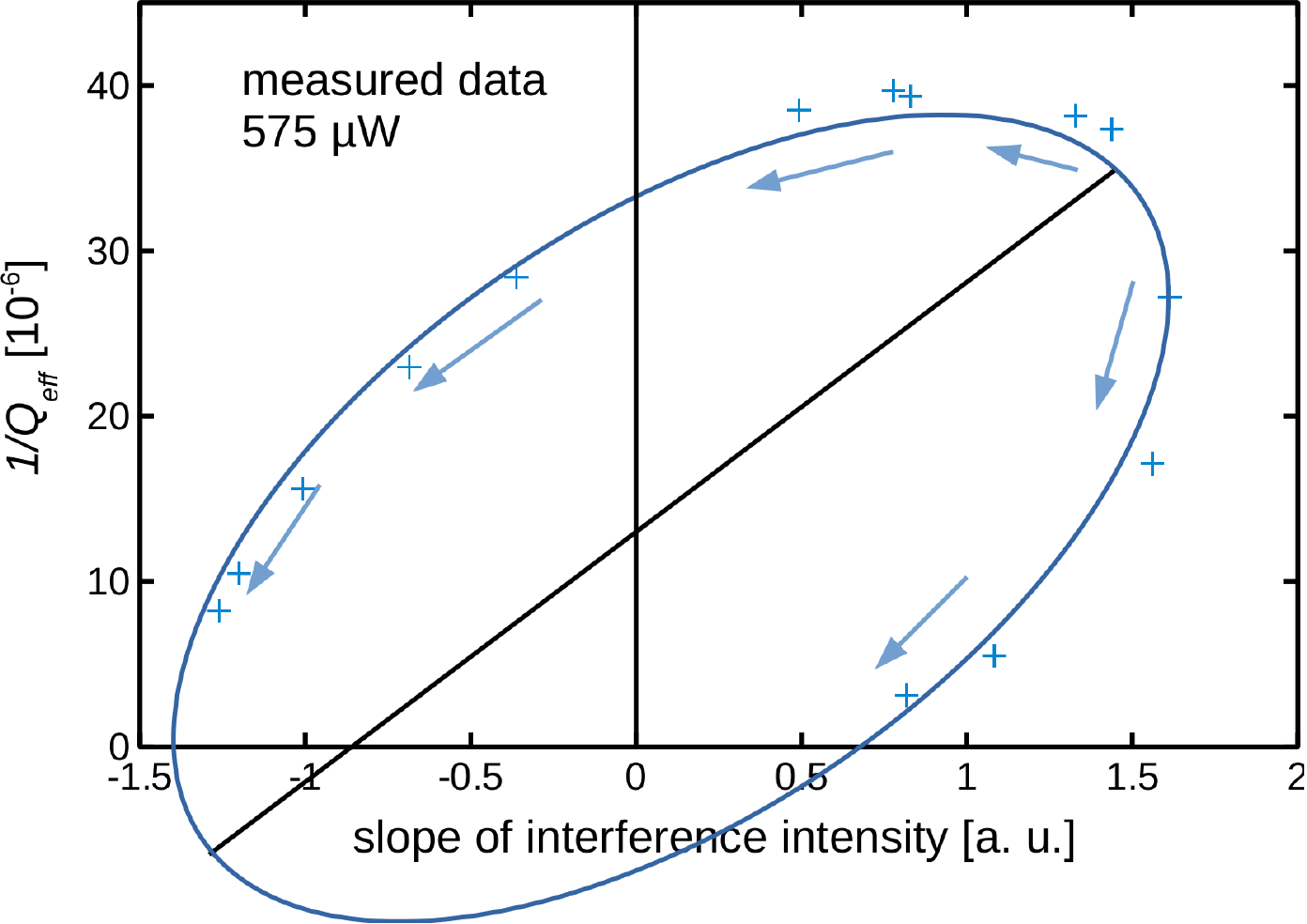}
\caption{\label{fig:lissajous} A Lissajous figure of the inverse of the effective quality factor is observed as a function of the slope of the interference intensity. The oval is a guide to the eye. Arrows indicate the path followed during the series of measurements.}
\end{figure}

For measuring the thermal noise and thermally excited self-oscillations of a cantilever in a Fabry-Perot cavity, we utilised our Omicron UHV low-temperature AFM equipped with a homodyne optical fibre interferometer detection system operated with a laser of wavelength 830\,nm. We use the original laser detection unit shipped with the set-up. The laser interferometer consists of a laser diode (Sharp LT015MD) coupled to a $2\times2$-fibre optic coupler (GOULD fibre optics 2581063) on input~1. A photodiode (Hamamatsu S5972) is coupled to input~2 and measures the interfering light reflected back from the cantilever as well as from the end of the bare fibre connected to output~1 of the fibre coupler. The second output of the fibre coupler is cut off, such that no back reflections reach the photodiode. For additional details, see also \cite{hug99p1}. The reflectivity of the fibre end is about $R_1=4$\%. We mounted a 225\,$\mu$m long, 28\,$\mu$m wide and 3\,$\mu$m thick Si PPP-MFMR cantilever (Nanosensors, Neuch\^atel, Switzerland) coated with a 30\,nm thick Al layer as reflection coating and cooled down the system to 10\,K. The cantilever had a free resonance frequency of 74.083\,kHz at 10\,K in UHV and a nominal spring constant of 2.8\,N/m. This value was provided by the manufacturer. Due to the reflection coating the cantilever reflects nearly $R_2=100$\% of the laser light. In this case we expect photothermal effects to dominate over radiation pressure \cite{marti92p1,flaeschner15p1}. Although upon reflection of the laser light only a small amount of energy is transmitted to the cantilever, this small amount is sufficient to induce a crucial bimetallic effect of the metal-coated Si cantilever. The finesse of the cavity \cite{svelto} was approximately $F=\pi(R_1 R_2)^{1/4}/(1-(R_1 R_2)^{1/2})\approx 1.8$. The working point of the interferometer was set in such a way that the slope $\mathrm{d}I/\mathrm{d}D$, where $I$ is the intensity and $D$ is the cantilever-fibre distance, was negative for all of the measurements except for the ones shown in Fig. \ref{fig:lissajous}.

\begin{figure*}
\includegraphics[width=0.9\linewidth]{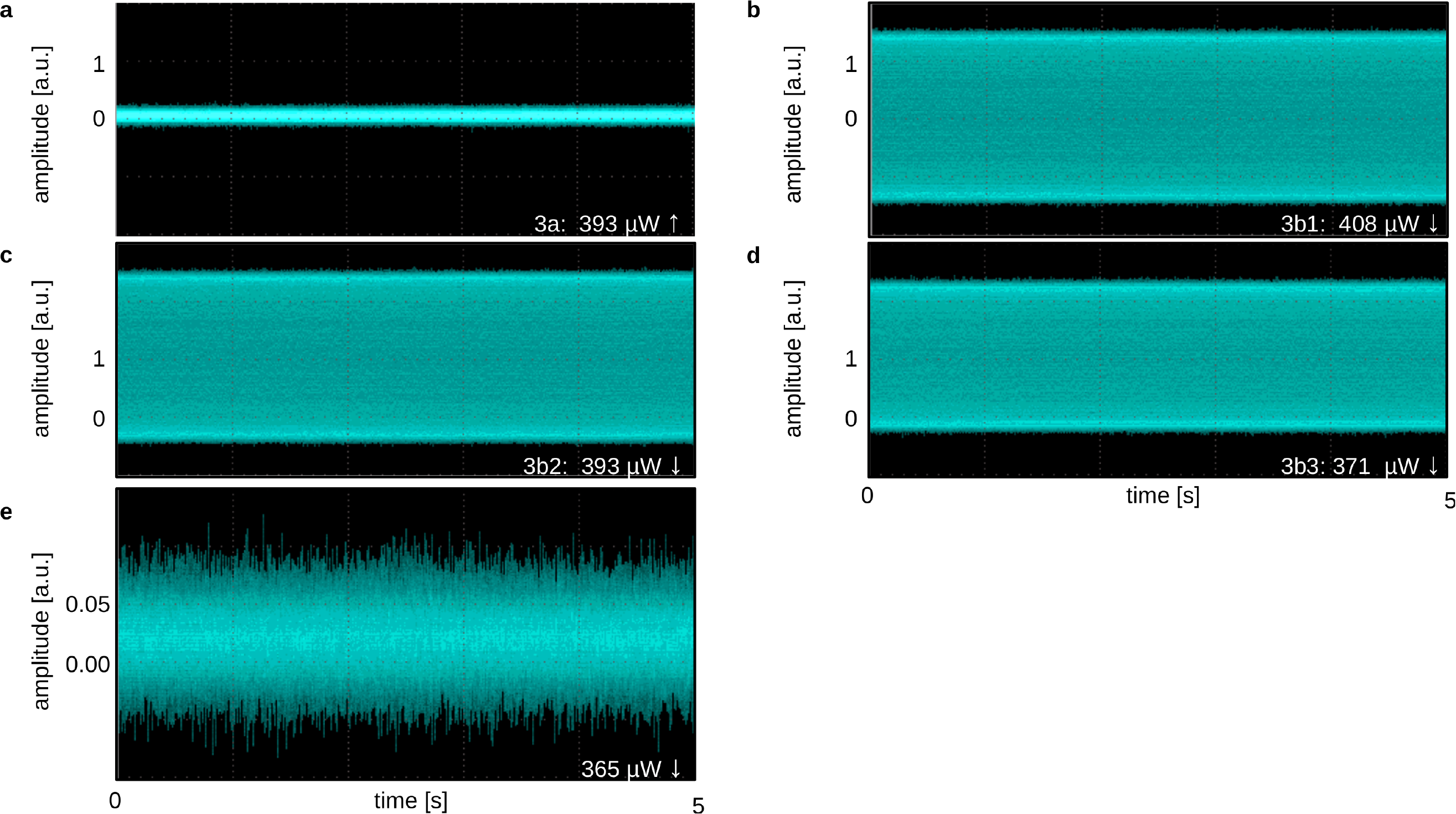}
\caption{\label{fig:timetraces} Examples of time traces. a) Time trace obtained on the lower branch of the amplitude hysteresis curve. b), c) and d) Time traces obtained on the upper branch of the amplitude hysteresis curve. Four of the corresponding spectra are shown in Fig. \ref{fig:spectra}. The positions where the time traces and spectra a) and b) to d) were taken are indicated in Figures \ref{fig:hystA} and \ref{fig:hystQ}. e) A spectrum with enlarged amplification of small amplitudes shows that small amplitude time traces show enhanced fluctuations.}
\end{figure*}

\subsection{\label{subsec:22}Intrinsic quality factor and measurements as a function of slope of interference intensity}

The intrinsic quality factor of the cantilever $Q_0=31 000$ was determined using the method described in Ref. \citenum{hoelscher09p1}: For these measurements the cantilever was driven mechanically and $Q_{\rm eff}$ was determined as a function of the slope of the interferometer. A Lissajous figure has been observed  (Fig. \ref{fig:lissajous}), because both the inverse of the quality factor and the slope of the interferometer are nonmonotonic, approximately sinusoidal functions of the cantilever position with respect to the interferometer fringe. This Lissajous figure differs from the ones previously observed: a straight line \cite{hoelscher09p1} and an eight-shaped Lissajous figure \cite{flaeschner15p1} have been reported previously. This means that the phase of the optomechanical effect is different from previous reports. This point will be studied in more detail.

\begin{figure}
\includegraphics[width=0.75\linewidth]{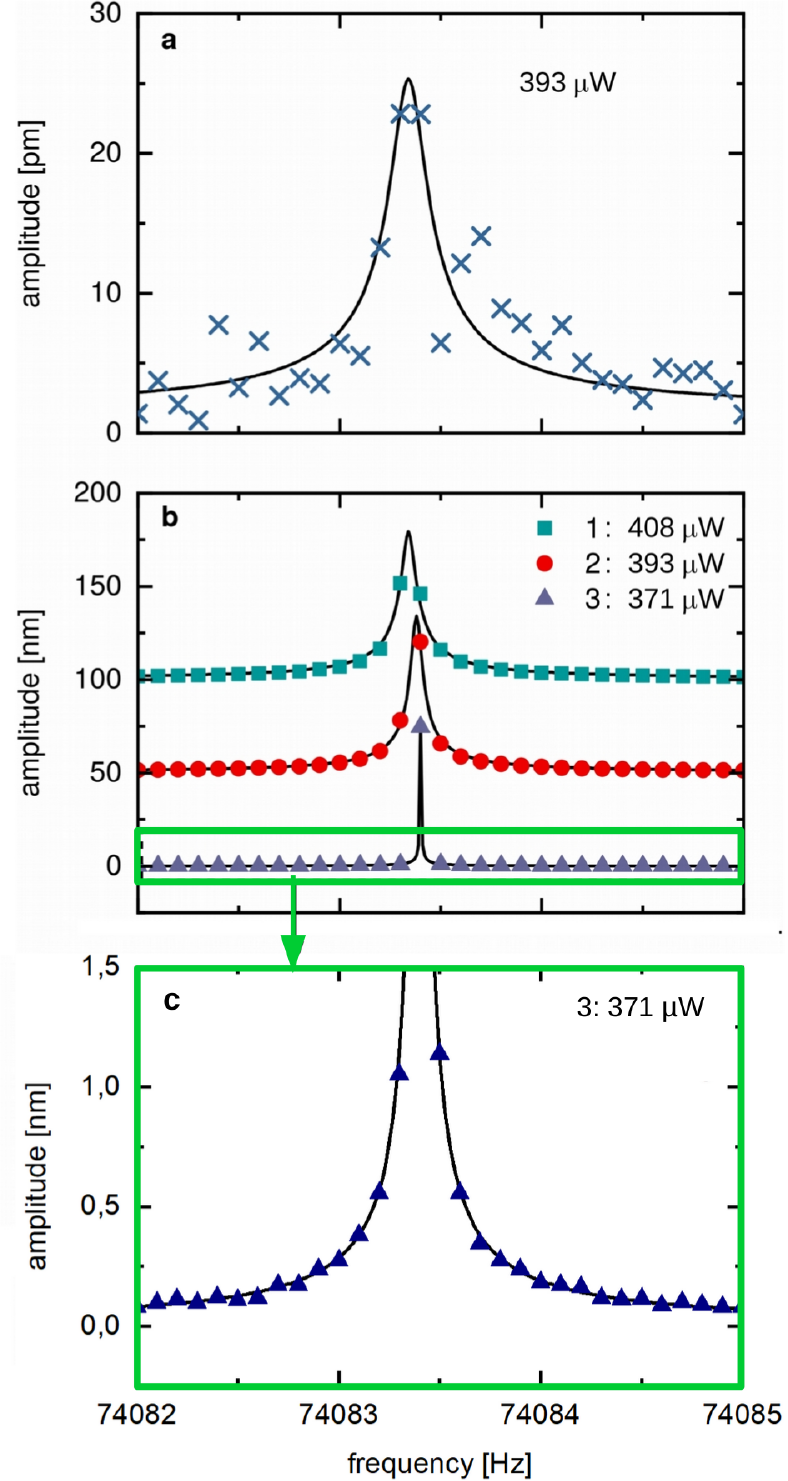}
\caption{\label{fig:spectra} Examples of thermal noise and thermally excited self-oscillation spectra obtained at different positions of the hysteresis curve. a) Spectrum obtained on the lower branch of the amplitude hysteresis curve. b) Spectra obtained on the upper branch of the amplitude hysteresis curve. To increase comparability, spectra 1 to 3 in b) have been plotted in the same graph with different offsets (1: 100\,nm, 2: 50\,nm, 3: 0\,nm). The positions where the spectra a) and 1 to 3 in b) were taken are indicated in Figures \ref{fig:hystA} and  \ref{fig:hystQ}. The shift of the resonance between spectra 1 to 3 in b) is 0.06 Hz at most as can be seen from the values in Table \ref{tab:fitvalues}. c) An enlarged area of spectrum 3 shows that more than ten nonzero data points contribute to the description of the spectrum.}
\end{figure}

\subsection{\label{subsec:23}Thermal noise and thermally excited self-oscillations measurements}

We measured the oscillation of the cantilever in the absence of mechanical driving while the laser power was set to a specific value using a digital oscilloscope (DL\,9240, Yokogawa, Japan) with a sampling rate of 250\,kSample/s for 10\,s. It is important to note that the maximal amplitude was always much smaller than the width of the interferometer fringe given by a quarter of the wavelength of the laser in order to avoid that the cantilever oscillation could surpass one interferometer fringe. The value of the laser power was changed manually after finishing the measurement to obtain the subsequent data point. The settling time prior to each measurement was at least 70\,s. This is reasonable for the data points marked in blue in Figures \ref{fig:hystA}a) and \ref{fig:hystQ}a), because the amplitude does not change in this region while $Q_{\rm eff}$ strongly increases (see Figures \ref{fig:hystA}a) and \ref{fig:hystQ}a)). If the settling time were insufficient, we would expect that increasing the settling time would lead to even larger values of $Q_{\rm eff}$. For the first measurement with a strongly reduced oscillation amplitude beyond the critical point of the hysteresis, the data point marked in green at 367 $\mu$W laser power on the black branch of Figures \ref{fig:hystA}a) and \ref{fig:hystQ}a), the settling time was increased to more than 3\,min~30\,s. If the measurement time had been enlarged even further we had expected additional uncertainty due to thermal drift. The interferometer signal was mixed (i.e., multiplied) with a frequency of 70\,kHz using a home-built multiplier based on the commercial component AD734. As a result, the interferometer signal oscillated at 4.083\,kHz = 74.083\,kHz – 70\,kHz. The high-frequency part at 144\,kHz was cut off by a low-pass filter. Typical data are shown in Fig. \ref{fig:spectra}. The data was fast Fourier-transformed using the digital oscilloscope. As a result we obtained a resonance spectrum of our cantilever, see Fig. \ref{fig:spectra}. We calculate the effective voltage in the frequency domain $V_{\rm eff}$ from the measured spectral density distribution $L$ in dBV via $V_{\rm eff} = 10 ^{L/20}$. Subsequently, we use a calibration spectrum with a known excitation ($V_{\rm pp} = 100$~mV) to determine a conversion factor, which allows the conversion of $V_{\rm eff}$ in the frequency domain to $V_{\rm pp}$ in the time domain. Then, we convert $V_{\rm pp}$ to a mechanical amplitude of the cantilever considering the interferometer’s slope and the sensitivity of the light detecting unit ($415$ nm / $13 V_{\rm pp}$). For spectra with a very low mechanical amplitude, we additionally applied a signal amplification of a factor of $10$ and we take it into account during the conversion.

In Fig. \ref{fig:spectra}, we represent the oscillation amplitude rather than the power spectral density. The power spectral density can be calculated from the oscillation amplitude. The representation in length units allows a direct comparison to interatomic distances needed for the operation of an AFM.

\subsection{\label{subsec:24}Curve fitting}

For fitting the spectra we used the well-know textbook formula for a harmonic oscillator\cite{Hoelscher2007}
\begin{equation}
A(f) = A_{\rm floor}+\frac{A_{\rm max}/Q_{\rm eff}}{\sqrt{\left(1-\left(f^2/f_{0}^2 \right)\right)^2 + \left( (1/Q_{\rm eff})(f/f_0)\right)^2} },
\label{eqn-res}
\end{equation}
\noindent
where $A_{\rm floor}$ is the noise floor of the measurement and $A_{\mbox{\scriptsize max}}$ is the maximal amplitude. We note that we expect deviations from this formula due to the nonlinearities. However, it will be shown below that the formula does describe the experimental data in a phenomenological way even in the presence of nonlinearities. We take the effective quality factor to be the value derived from a fit of the spectrum to this formula in all cases.

For spectra with a relatively small effective $Q$-factors and oscillation amplitudes, i.e. for the spectra of the lower branch of the hysteresis curve, the resonance peak is described by a few data points with considerable noise. In these cases, the frequency resolution is sufficiently good to derive all four variables ($A_{\mbox{\scriptsize max}}$, $f_0$, $Q$ and $A_{\rm floor}$) from a simple fit, but the values show a considerable mathematical uncertainty, as shown in table \ref{tab:fitvalues}. The results are reliable, because all four variables remain similar for all spectra of this branch, in particular the resonance frequency remains nearly constant under the conditions used here. As one can see from the spectrum shown in Fig.\,\ref{fig:spectra}a), the noise floor is an important parameter. For these low-$Q$ spectra the noise floor was therefore taken into account for fitting. The oscillation amplitude is only about $0.01$ nm to $0.1$ nm for low-$Q$ spectra of the lower amplitude branch such as the one shown in Fig. \ref{fig:spectra}a), much smaller than in usual AFM measurements. The mathematical uncertainty in the fit translates into fluctuations of the lower branch of the hysteresis (Figures \ref{fig:hystA} and \ref{fig:hystQ}). Instead of adapting the measurement set-up for an enhanced precision in this regime, we have chosen to focus mainly on the high oscillation amplitude branch.

For the spectra with large oscillation amplitudes, i.e. for the spectra of the upper branch of the hysteresis curve, (see Fig. \ref{fig:spectra}b), the oscillation amplitude noise floor level was neglected in these fits. This includes the spectra with particularly large $Q$-factors. For spectra with large $Q$-factors the frequency resolution is insufficient due to the extremely tiny width of the resonance peak. This applies in particular to the spectrum labelled 3 in Fig.\, \ref{fig:spectra}b). Nevertheless the spectrum is described by more than ten nonzero data points as shown in Fig. \ref{fig:spectra}c).

In order to describe also these spectra accurately, we used several fitting procedures. All of them showed qualitatively similar results: 1. In our first approach, we first determined $A_{\rm max}$ from the value obtained from the previous spectrum and from the maximal amplitude shown in the spectrum. Then $f_0$ and $Q_{\rm eff}$ were obtained from a simultaneous fit. After that $f_0$ was set constant and $A_{\rm max}$ and $Q_{\rm eff}$ were obtained from a simultaneous fit. 2. In a second attempt, we obtained $A_{\rm max}$ from the envelope of the oscillator signal in the time domain, which we then used as a mathematically fixed value for $A_{\rm max}$ of the fit formula (Eq. (\ref{eqn-res})) in the frequency domain. We then obtained $f_0$ and $Q_{\rm eff}$ from automated fitting. 3. In an effort to become even more accurate we used a third approach. The results of this third approach are shown in the Figures. First, we determined the cantilever’s amplitude from the time domain and mathematically fixed $A_{\rm max}$ as in the second approach. Then, we manually determined the resonance frequency $f_0$ by minimising the error of $Q_{\rm eff}$ of the automated fitting routine. Note that the accuracy given for the resonance frequency applies only to its relative value but not to the absolute value. 

In this way, we reduced the number of free variables for fitting from four to one. Starting from the cantilever oscillation amplitude measured in the time domain enabled us to fit the resonance frequency with a precision below the frequency resolution of the interferometer. 

Nonetheless, fitting led to uncertainties for the values of $A_{\rm max}$ and $Q_{\rm eff}$ of up to 14 percent. In spite of these comparably large uncertainties we obtain significant results for the hysteresis of $A_{\rm max}$ and $Q_{\rm eff}$. In particular the results for $Q_{\rm eff}$ for the data point with the highest value are in line with the other values in the inset of Figure \ref{fig:hystQ}.

\begin{table*}
\caption{\label{tab:fitvalues}The table contains the values through the described fitting procedure together with the respective standard deviation obtained through curve fitting. Values where no standard deviation is given were kept fixed during fitting and were obtained as described in the text.}
\begin{ruledtabular}
\begin{tabular}{ccccc}
Fig. & $A_{\mbox{\scriptsize max}}$ [nm] & $f_0$ [Hz] &$Q_{\mbox{\scriptsize eff}}$&$A_{\rm floor}$ [nm]\\
\hline
\ref{fig:spectra}a & $2.4 \times 10^{-2}\pm 1.4 \times 10^{-3}$ & $74083.34$ &$4.3 \times 10^5 \pm 4.3\times 10^4$&$1.4 \times 10^{-3} \pm 7.9 \times 10^{-5}$\\
\ref{fig:spectra}b1 & $79.4$ & $74083.34$ &$1.1 \times 10^6 \pm 8.2 \times 10^3$&$0$\\
\ref{fig:spectra}b2 & $84.2$ & $74083.38$ &$1.4 \times 10^6 \pm 8.6 \times 10^3$&$0$\\
\ref{fig:spectra}b3, \ref{fig:spectra}c & $78.9$ & $74083.40$ &$2.6 \times 10^7 \pm 3.5 \times 10^4$&$0$\\
\end{tabular}
\end{ruledtabular}
\end{table*}

\section{\label{sec:level3}Experimental results}

\begin{figure}
\includegraphics[width=0.9\linewidth]{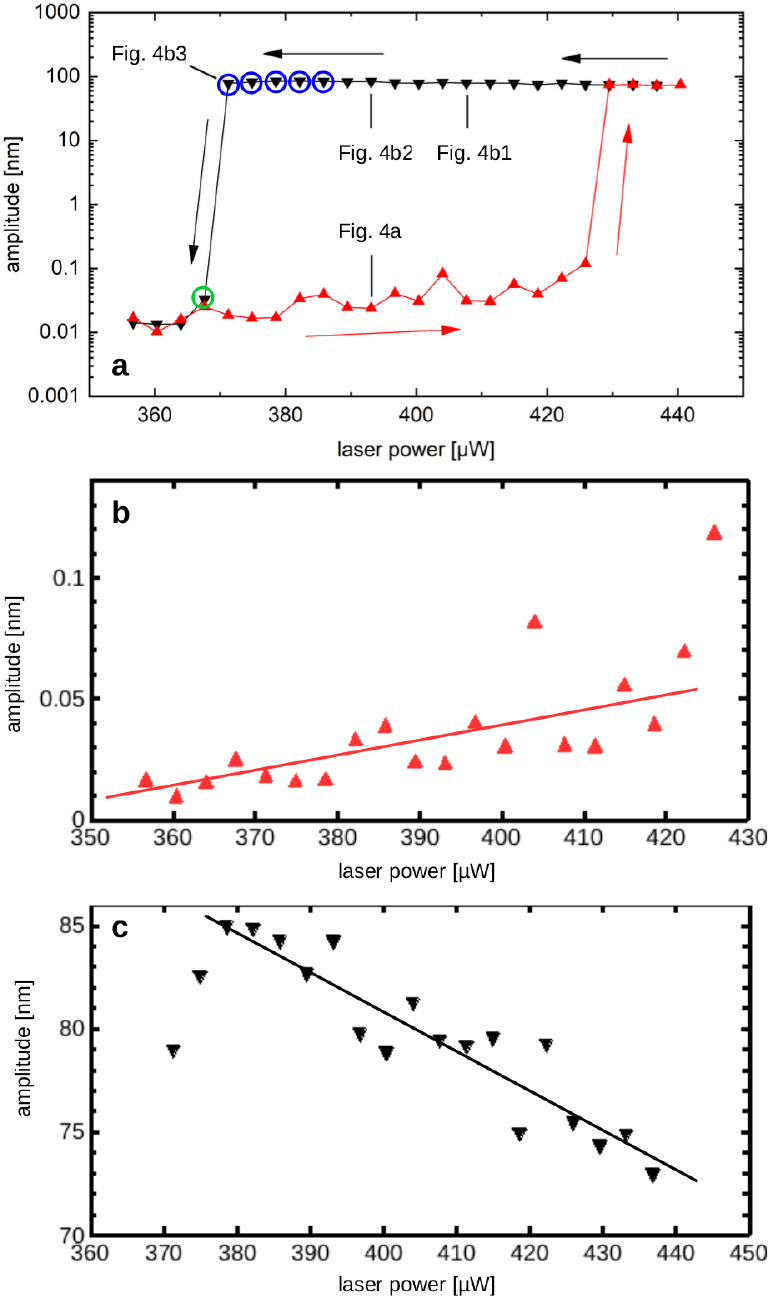}
\caption{\label{fig:hystA} a) Logarithmic plot of the cantilever's amplitude as a function of the laser power. This amplitude obtained from the thermal noise spectra or from time traces shows hysteresis during increase and decrease of laser power. The data points obtained from the fits shown in Fig.\,\ref{fig:spectra} are labelled. The last five data points with a large oscillation amplitude on the black branch are marked in blue, the first data point with a strongly reduced oscillation amplitude is marked in green. Linear plot of the amplitude as a function of laser power for b) the lower branch and c) the upper branch of the hysteresis curve. The lines serve as a guide to the eye.}
\end{figure}

\begin{figure}
\includegraphics[width=\linewidth]{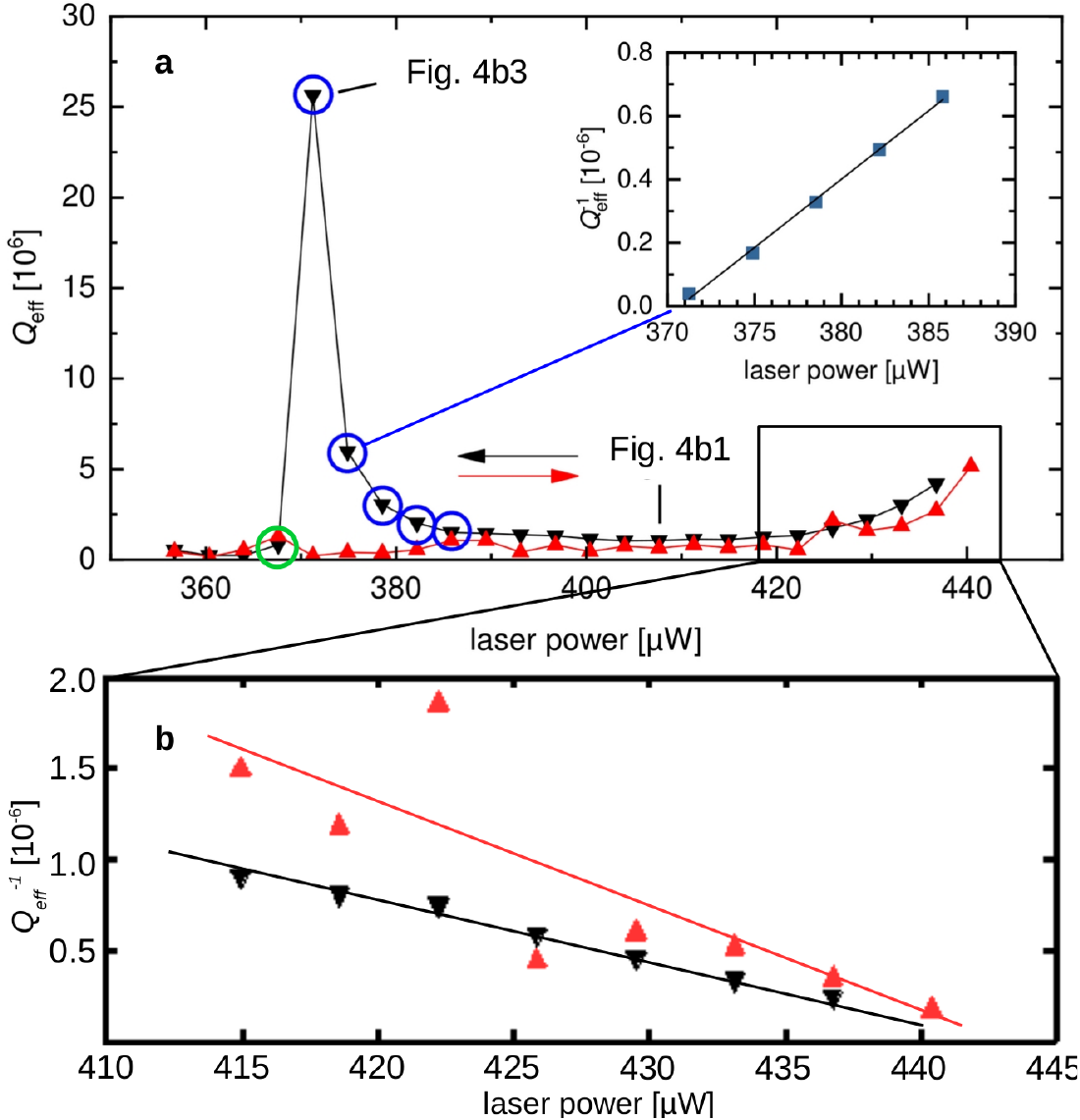}
\caption{\label{fig:hystQ} a) Plot of the $Q$-factor as function of the laser power reveals a hysteresis, too. Prominent is a strong increase of the $Q_{\rm eff}$ factor near the point where the amplitude falls back to the lower branch of the hysteresis. Points showing signs of this increase are highlighted with blue circles. The first data point with a strongly reduced oscillation amplitude also shows a strongly reduced $Q_{\rm eff}$ factor and is marked in green. This increase is explained by the occurrence of a critical point related to a Hopf bifurcation near the point of maximal $Q_{\rm eff}$ factor. Inset: Linear dependence of the inverse effective $Q_{\rm eff}$ factor on laser power for the region where the strong increase of $Q_{\rm eff}$ is observed (points marked in blue). b) Near the point where the self-oscillation starts, $Q_{\rm eff}^{-1}$ shows the expected linear behavior as a function of laser power as described by equation \ref{eqn-effQ}. The lines serve as a guide to the eye. On the low amplitude branch, the small oscillation amplitude leads to additional fluctuations in the region where $Q_{\rm eff}$ is small near a laser power of 425 $\mu$W.}
\end{figure}

To summarize, time traces from the measurements are shown in Fig. \ref{fig:timetraces}. The spectra taken for different laser powers are shown in Fig.\,\ref{fig:spectra}. The fitted values of $A, Q_{\rm eff}, f_0$ and $f_{\rm floor}$ obtained from fitting the spectra shown in Fig.\,\ref{fig:spectra} are listed in table\, \ref{tab:fitvalues}. The oscillation amplitude and the effective quality factor of the cantilever obtained by fitting all data are shown in Fig.\,\ref{fig:hystA} and \ref{fig:hystQ}.

As we vary the laser power between 350 and 440\,$\mu$W, we observe that the oscillation amplitude of the cantilever indeed shows the expected hysteresis (Fig. \ref{fig:hystA}). With increasing laser power, the oscillation amplitude first remains small, on the order of 0.01 to 0.1 nm. It increases only slightly with laser power, Fig. \ref{fig:hystA}b). For larger laser powers above $426 \mu$W, a critical point is reached where self-oscillation sets in and the oscillation amplitude jumps to a large value of around $80$\,nm. As the laser power increases above the critical value, the $Q_{\mbox{\scriptsize eff}}$ factor becomes negative in a non-steady state that has not been captured, and the amplitude increases until it reaches a new steady state where nonlinearities stop it from increasing further and cause the effective $Q_{\rm eff}$ factor to become positive again. When the laser power is subsequently decreased the oscillation amplitude remains at this large value (around 80 nm) until again a critical point is reached where it jumps back to a small value. On the upper branch of the hysteresis curve, the oscillation amplitude slightly increases with decreasing laser power, Fig. \ref{fig:hystA}c).

For the effective quality factor, near both the upper and lower critical points of the hysteresis, an increase of $Q_{\rm eff}$ is observed (Fig. \ref{fig:hystQ}a). The quality factor of the cantilever remains below $6 \times 10^6$ for most of the hysteresis with the exception of an intense peak near the point where the oscillation amplitude falls back to the lower branch for decreasing laser power. This peak cannot be explained by noise of the measurement, because in its maximum it exceeds by far the noise floor since the $Q_{\rm eff}$-factor is more than four times larger than anywhere in the other data. Also there are at least four data points that show an increasing $Q_{\rm eff}$-factor as the laser power approaches the critical point. Since the $Q_{\rm eff}$-factor is increased further by a factor of 18 with respect to the part of the hysteresis where the $Q_{\rm eff}$-factor is high (3 in Fig. \ref{fig:spectra}b) compared with 2 in Fig. \ref{fig:spectra}b)), we expect the thermal noise of the cantilever calculated from Eq. \ref{eqn-noise} to be around four times smaller in this area. If we compare with the part of the hysteresis where the $Q_{\rm eff}$-factor is low (3 in Fig. \ref{fig:spectra}b) compared with Fig. \ref{fig:spectra}a)), we even obtain a factor of 60 for the enhancement of the $Q_{\rm eff}$-factor, which translates into a factor of around eight for the noise reduction. By tuning the laser power even better to the optimal point, we believe that the $Q_{\rm eff}$-factor could be increased even further.

The intense increase of the $Q_{\rm eff}$-factor occurs on the branch where the oscillation amplitude is large (upper branch) at the position near the jump to the lower branch. This is also the critical point described by the predicted subcritical Hopf bifurcation \cite{zaitsev12p1}. For this prediction the equation of motion of the cantilever oscillation is described including higher-order nonlinear force terms as had been used before for carbon nanotube and graphene oscillators \cite{eichler11p1}. An effective $Q_{\rm eff}$-factor depending on the laser intensity has been defined previously using a linear equation of motion for the cantilever and a sinusoidal oscillation with a modified oscillation frequency $\omega_m$ and phase \cite{hoelscher09p1}. In this equation of motion, nonlinearities arise directly from the photothermal force occurring exponentially with a time delay and introducing nonlinearities due to the exponential dependence on time. From this model, we have obtained previously \cite{hoelscher09p1} the modified resonance curve, and have defined an effective quality factor $Q_{\rm eff}$. In this model, $Q_{\rm eff}$ depends on laser intensity as

\begin{equation}
\frac{1}{Q_{\rm eff}}=\frac{1}{Q_0}-\frac{c_{ph}}{c_z}\frac{\omega_0 \tau}{\left(1+\tau^2\omega_m^2\right)}
\label{eqn-effQ}
\end{equation}
\noindent
where $c_{\rm z}$ is the cantilever spring constant, $c_{\rm ph} =\partial F_{\rm ph}/\partial d$ is the optical spring constant (where $d$ is the cavity length) and $\tau$ is a constant describing the time delay of the photothermal force. The optical spring constant increases linearly with the intensity of the photothermal force. Upon starting at small values and increasing the laser power, the optical spring constant increases up to the point where $Q_{\rm eff}^{-1}$ becomes first zero and then negative. At this point, the damping of the cantilever turns into a driving force, and the cantilever starts to self-oscillate. It is now in a different state concerning the nonlinearities induced by the photothermal force where the oscillation amplitude is larger. In this new state, $Q_{\rm eff}^{-1}$ has a new positive value. Upon decreasing the laser power, $Q_{\rm eff}^{-1}$ again becomes smaller and approaches zero. At another critical point, the solution with the large amplitude becomes unstable, the excitation turns into a damping again and the oscillation amplitude falls back to the lower branch.

\begin{figure}[b]
\includegraphics[width=0.8\linewidth]{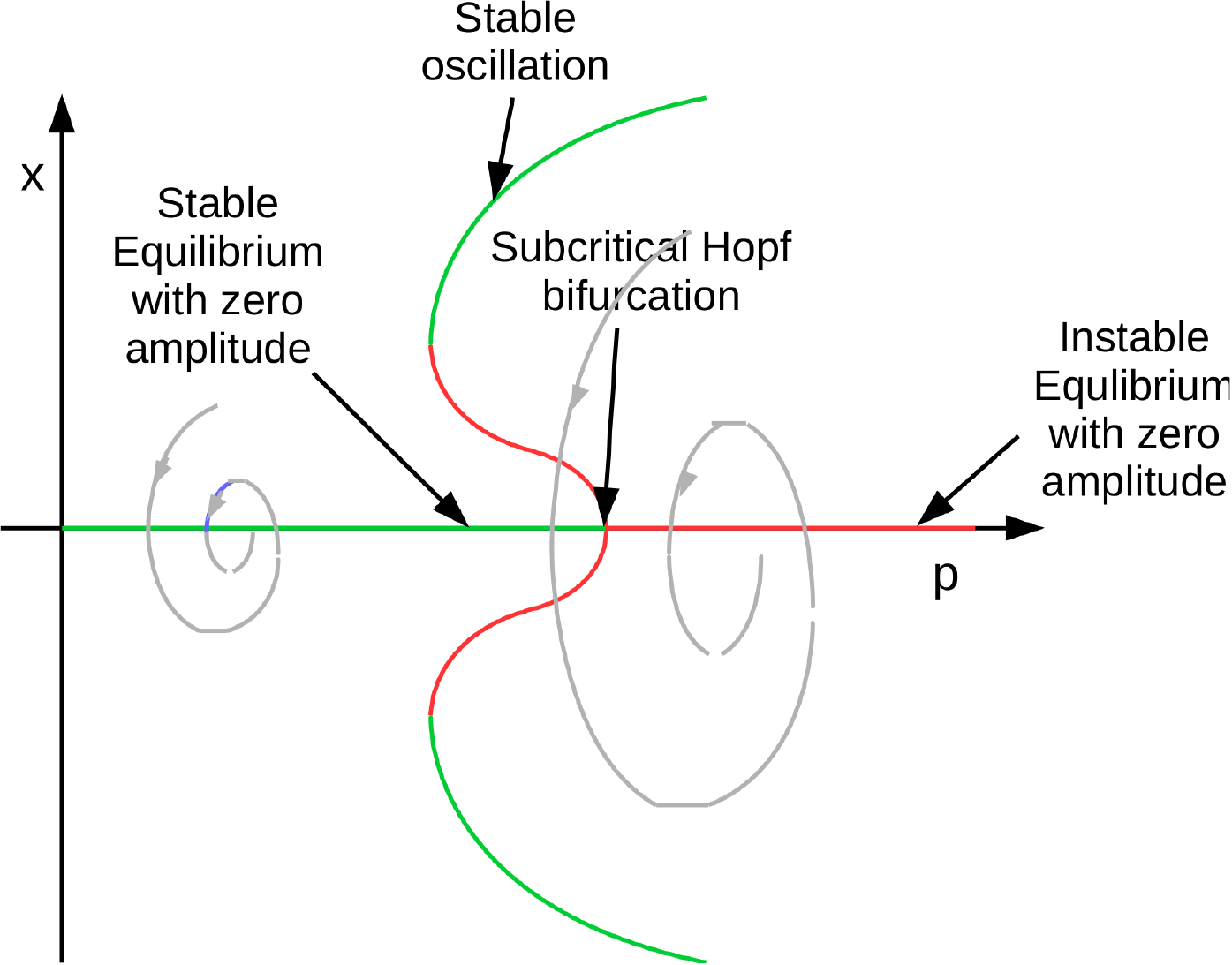}
\caption{\label{fig:vorst} Schematic description of oscillation depending on the parameter $p$ that represents laser intensity for the mathematical experiments. For small $p$ the oscillations remain small and decay, as the stable solution shows zero amplitude. Upon increasing $p$, this solution becomes instable and a stable oscillation branch occurs. If $p$ is decreased from this large value, the point where the branch with stable zero amplitude is reached differs from the point where it first occurred due to non-linearities.}
\end{figure}

The model described above had been developped to describe the situation within one interferometer-fringe. To check experimentally whether Eq. \ref{eqn-effQ} describes the experimental situation also near the critical points of the hysteresis independently of the role of the nonlinearities, we have plotted $Q_{\rm eff}^{-1}$ as a function of laser power for the four points where the increase was observed in Fig.\,\ref{fig:hystQ} (points marked in blue). Indeed, a linear dependence on laser power is observed with $Q_{\rm eff}^{-1}$ approaching zero for the critical point. A similar linear dependence of $Q_{\rm eff}^{-1}$ on the laser power is also observed at the upper critical point of the hysteresis (Fig.\,\ref{fig:hystQ}b)).

Inspite of the fluctuations it is remarkable that both curves show an upturn or a downturn in the region where the self-oscillations starts or ends. This behaviour is drawn schematically in Fig.\,\ref{fig:SetUp} and Fig.\,\ref{fig:vorst} (green curve) and the downturn is also observed in simulations in Fig.\,\ref{fig:p050}.

\section{\label{sec:level4}Equation of motion of the cantilever and numerical experiments}

\subsection{General form of the equation of motion}

The equation of motion of the cantilever is the well-known one of a driven damped linear mechanical oscillator. We now additionally introduce the photothermal $F_{\mbox{\scriptsize photo}}(x)$ and radiation pressure force $F_{\mbox{\scriptsize rad}}(x)$.

\begin{equation}
    \ddot{x}+\frac{\omega_0}{Q_0}\dot{x}+\omega_m^2\cdot x=2f_m\cos\left(\omega_0 t+\phi_0\right)+\frac{1}{m}\left[ F_{\mbox{\scriptsize rad}}(x)+F_{\mbox{\scriptsize photo}}(x)\right],
    \label{eqn-mottotal}
\end{equation}
\noindent
where $x$ is the position of the mechanical oscillator with respect to the interferometer fringe, $\omega_0 = 2\pi f_0$ is the resonance frequency of the cantilever, $\omega_m$ is the resonance frequency modified by the interaction with the interferometer, and $f_m$ is the external driving force. In the original AFM set-up, non-linearities are negligible. The interaction with the light field introduces nonlinearities, because it is represented by an exponential function with a time delay. One usually motivates the existence of nonlinearities from the dependence of the radiation pressure force on $x$. However, here we operate in a regime where the photothermal force is large compared to the radiation pressure force. We expect and have shown \cite{Hoelscher2007} that in our case the photothermal force is a nearly sinusoidal function of $x$, introducing even ($\sim \cos x$) or uneven ($\sim \sin x$) nonlinearities depending on its phase related to its time delay. For symmetry reasons terms of uneven order are grouped with Hooke's laws term, whereas terms of even order are grouped with the damping term. The idea is that the main part of the damping is always opposed to the direction of the velocity and that the potential is mainly symmetric with respect to the rest position.

The most general form of the equation of motion including all possible nonlinearities is

\begin{eqnarray}
    \ddot{x}&+&\left(\frac{\omega_0}{Q_0}+\gamma_3 x^2+\gamma_5 x^4+\ldots\right)\dot{x}+\left(\omega_m^2+\alpha_3 x^2+\alpha_5 x^4+\ldots\right)x\nonumber\\
    &=&2f_m\cos\left(\omega_0 t+\phi_0\right),
    \label{eqn-mottotal1}
\end{eqnarray}
\noindent
where $\alpha_3$, $\alpha_5$, $\gamma_3$ and $\gamma_5$ are nonlinear coefficients. The nonlinear terms are now designed to include all nonlinear effects that occur in the system.

Here, we are particularly interested in the variation of the effective quality factor that is mainly responsible for the hysteresis we observe as described above. A variable $Q_{\rm eff}^{-1}$ can be understood as a variable damping coefficient. Although nonlinear effects occur in both, potential (or force) and damping, here, we focus on nonlinear effects in the damping coefficient, because we observe its variation directly in the experiment. In addition, since we place the cantilever at the center of the interferometer fringe, and we know that the photothermal force drives the cantilever, we expect that the important effects are related to the terms of even powers in the damping coefficient.

\subsection{Numerical experiments}

To show that nonlinear damping can indeed describe the experiment, we have performed numerical experiments using several simplified forms of the equation of motion (Eq.\,\ref{eqn-mottotal1}). From Eq. \ref{eqn-effQ} and its experimental validation, we know that $Q_{\rm eff}^{-1}$ varies linearly with laser power near the bifurcation points of the hysteresis, because the optical spring constant varies linearly with laser power. For the numerical experiments, this dependence on laser power is described by a parameter $p$, taken to be proportional to the laser power. At sufficiently large laser intensity $Q_{\rm eff}^{-1}$ becomes zero, causing self-oscillations limited in amplitude by the nonlinearities of the system. Upon decreasing the laser intensity there is another point where $Q_{\rm eff}^{-1}$ becomes zero.

\begin{figure}
\includegraphics[width=\linewidth]{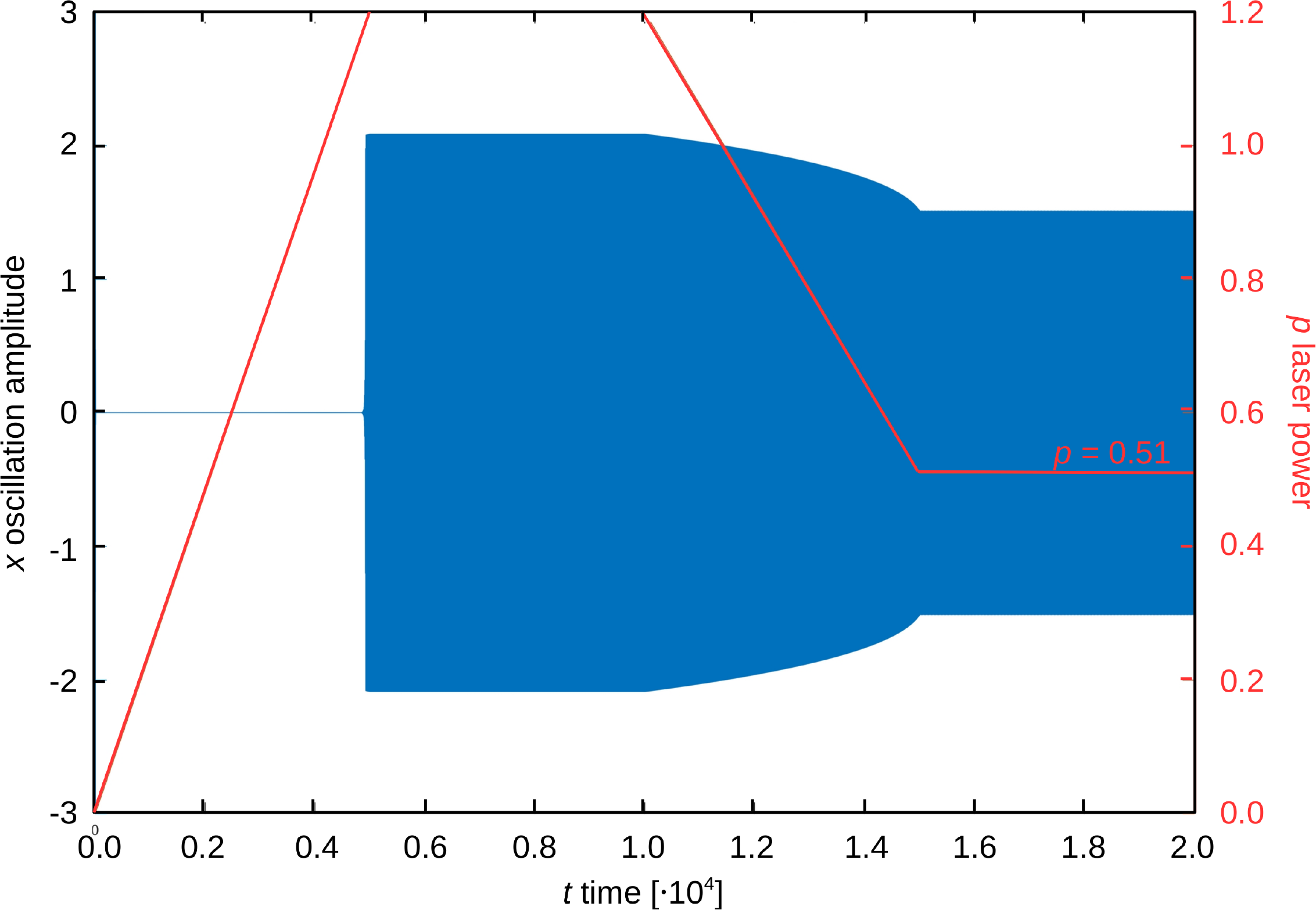}
\caption{\label{fig:p051} In the numerical experiments, we start with $p=0$ and increase $p$ up to values larger than one. An oscillation is observed starting from $p\approx 1.1$. Since the period of the oscillation is much smaller than the resolution of the graph, the graph shows the envelope of the oscillation. Then $p$ is decreased again. For $p=0.51$ the oscillation remains. The axes of the graph show absolute numbers.}
\end{figure}

\begin{figure}
\includegraphics[width=\linewidth]{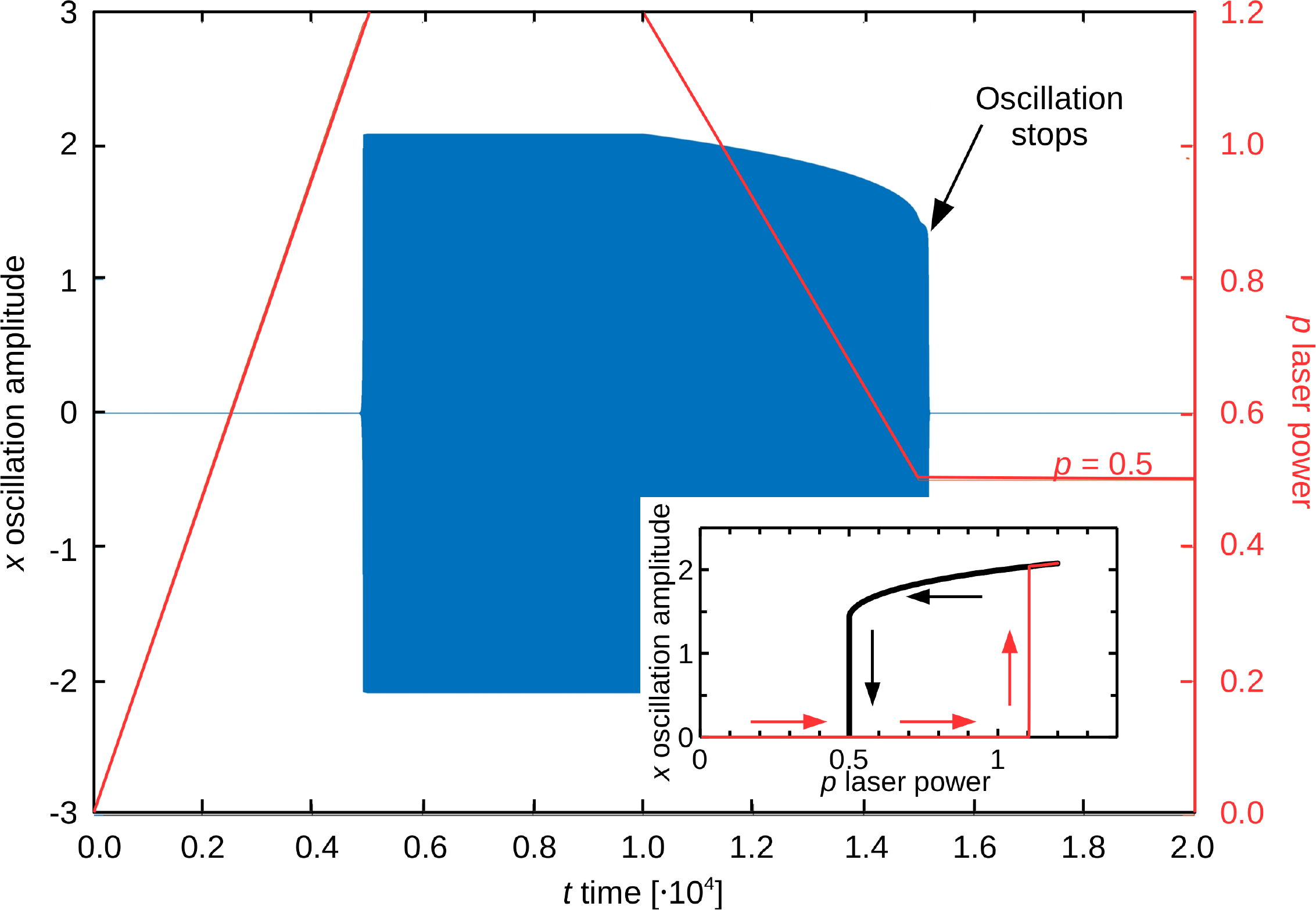}
\caption{\label{fig:p050} Same numerical experiment as Fig. \ref{fig:p051}. Upon decrease, $p$ is now set to 0.50 and the oscillation amplitude decreases until it reaches zero amplitude. Inset: The amplitude data is shown as a function of laser power to demonstrate its hysteretic behavior.}
\end{figure}

In the simplest equation of motion containing nonlinear damping (van der Pol oscillator, see also \cite{vogel03p1}) the coefficient $\gamma_5$ and all higher order coefficients are set to zero

\begin{equation}
    \ddot{x}+\left(\frac{\omega_0}{Q_0}(1-p+\tilde{\gamma_3} x^2)\right)\dot{x}+\omega_m^2x=2f_m\cos\left(\omega_0 t+\phi_0\right).
    \label{eqn-mot2}
\end{equation}
\noindent
 However, no hysteresis can occur ignoring the nonlinear terms $\gamma_5$ or higher, because in all cases a supercritical Hopf bifurcation is obtained. This is related to the fact that a parabola shows only one extremal point. We, therefore, keep the nonlinear coefficient $\gamma_5$

\begin{equation}
    \ddot{x}+\left(\frac{\omega_0}{Q_0}\left(1-p+\tilde{\gamma_3} x^2+\tilde{\gamma_5} x^4\right)\right)\dot{x}+\omega_m^2x=2f_m\cos\left(\omega_0 t+\phi_0\right),
    \label{eqn-mot4}
\end{equation}
\noindent
and find that it is possible to obtain hysteretic behaviour using this equation, see Fig. \ref{fig:vorst} for a schematic description of the overall results. This equation has a subcritical Hopf-bifurcation for $p = 1$ which is related to the shape of the fourth order polynom with two minima and one maximum as is reflected in the W-shaped curve observed when one turns Fig. \ref{fig:vorst} by 90 degrees. The subcritical Hopf-bifurcation occurs for $\tilde{\gamma_3}<0$ and for $\tilde{\gamma_5}>0$. For the numerical simulations we have used $\tilde{\gamma_3}=-2$, $\tilde{\gamma_5}=1$, $\omega_0=1$ and $\sigma_0 =0.1$ and a very small driving force of $f_m = 10^{-4}$. We started the simulation with random values and have chosen a small step size such that the calculation lasted for a long time. As a function of time the parameter $p$ was varied. The value of $p$ at a given point in time is shown in red in the figures \ref{fig:p051} and \ref{fig:p050}.

Starting with $p = 0$, $p$ was increased up to $p = 1.2$ (a value larger than $1$), see Fig. \ref{fig:p051}. The oscillation of $x$ is represented in blue. At about $p = 1.1$ we observe the bifurcation and a stable oscillation is formed. The maximal value of $x$ assumes a value of 2 at $t=0.5 \times 10^{4}$, the individual oscillation is not resolved in this figure. When $p$ is reduced slowly, the oscillation remains stable down to $p = 0.51$ and assumes an oscillation amplitude of about $1.5$ at $t=1.5 \times 10^{4}$, see Fig. \ref{fig:p051}. If $p$ is reduced further (Fig. \ref{fig:p050}) the oscillation stops and the system assumes again the stationary state where $x = 0$ at all times starting from $t=1.5 \times 10^{4}$. The hysteresis resembles is represented in the same way as for the experiment (Fig. \ref{fig:hystA}) in the inset of Fig. \ref{fig:p050}.

The model intends to show how the linear dependence of $Q_{\rm eff}^{-1}$ on laser power (Equation \ref{eqn-effQ}) can be directly incorporated into the equation of motion of the cantilever: When $p$ equals 1 or larger values, the effective damping is negative for most of the oscillation cycle and the oscillation starts. When $p$ is reduced to zero, the damping term is positive during the full oscillation cycle and the oscillation is damped.

Indeed, including 4th order terms in the damping coefficient allows to explain fully both the observed hysteresis and the linear variation of $Q_{\rm eff}^{-1}$ near the bifurcation points of the hysteresis. In the future, we plan to extract the nonlinear coefficients from the experiment. Preliminary results show that the sign and the order of magnitude of $\tilde{\gamma_3}$ and $\tilde{\gamma_5}$ agree with the experiment.

\section{\label{sec:level5}Discussion}

In our model described in Eq. \ref{eqn-mot4}, the damping depends on oscillation amplitude, and a larger oscillation amplitude causes a larger damping. Preliminary measurements where the cantilever was driven externally, show that the oscillation amplitude indeed depends in a nonlinear way on the driving amplitude: The oscillation amplitude rises less than expected from a linear behavior. The increase of the oscillation amplitude becomes smaller with increasing driving force as described in our model. This point will be checked in more detail in additional experiments. From the model we expect that there should be a dependence of the oscillation amplitude on laser power. Such a dependence is indeed observed in Fig. \ref{fig:hystA}b) and c), but it is small. Also for the parameters used for the numerical experiments, a large change of $p$ from $1.2$ to $0.51$ leads only to a small change in oscillation amplitude (Fig. \ref{fig:p051}).

The physics of the variation of $Q_{\rm eff}^{-1}$ with the laser power and variable damping can be understood from literature, where the photothermal force has been described as having two components, a mechanical frequency shift due to heating, and a force depending on the temperature difference between the cantilever and its environment \cite{zaitsev11p1}, $\Delta T$. Our results suggest that in our case, the force due to the mechanical frequency shift due to heating is negligible. Mainly for the case of the radiation pressure force, the Physics of the forces has been understood from the non-sinusoidal shape of the force as a function of position $x$ for cavities with a high finesse. For the photothermal force, this relation should be investigated in more detail.

We expect that the nonlinear terms $\gamma_3$ and $\gamma_5$ could additionally depend on the laser power. This could be added to the model. Here, we have assumed that the nonlinear damping terms result from the photothermal and the radiation pressure force or both. We consider mainly the effect on the effective quality factor and on the oscillation amplitude and their noise or self-oscillations. We did not consider the effect of force noise on the cantilever motion.

The largest values we measure for $Q_{\rm eff}$ could be compared to the largest known intrinsic $Q$-factors, which are on the order of 5 million \cite{moser14p1, adiga12p1, poot12p1} and to the largest $Q$-factors induced by back-action, 100 million \cite{rossi18p1}. Using large effective $Q$-factors in scanning force microscopy generally represents an experimental problem: The frequency precision of the phase-locked loop system used for excitation and detection must be good enough to be able to fine tune the excitation frequency to the resonance frequency of the cantilever. The feed-back loop that keeps the oscillation amplitude constant must be extremely fast. Also the excitation amplitude must be small. Therefore, there have been proposals to reduce the $Q$ factor rather than enhancing it \cite{dagdeviren16p1}. Scanning force microscopy measurements with extremely large $Q$-factors have been successfully performed previously using a home-built phase-locked loop system \cite{hoffmann07p1}. In these experiments, an effective quality factor of 2 million has been observed in the regime of driven oscillations away from the Hopf bifurcation. Near the critical point located at the edge of the self-oscillation hysteresis, small deviations could cause large shifts in its oscillation properties. However, the large effective $Q$-factor also mean that the system remains in its oscillation state for a long time and that the oscillation is robust. It has been experimentally confirmed that large $Q$-factors are advantageous for low thermal noise \cite{luebbe13p1}.

Since the cantilever is driven by the interaction with the laser light, one could compare the cantilever-interferometer system with a self-excited system using electronic feedback self-excitation. In such systems, a modified (apparent) $Q$-factor differing from the original one is observed. However, this modification of the apparent $Q$-factor does not lead to changes of the force sensitivity \cite{albrecht91p1, flaeschner15p1}. In such linear systems, the increase of the apparent $Q$-factor is explained simply by an increase of the oscillation amplitude of the thermal noise \cite{albrecht91p1}. Here, in contrast we consider an oscillator with nonlinear damping and other additional nonlinear contributions and define the effective $Q$-factor by analogy to the linear system. Indeed, we observe that the quality factor is larger on the upper branch of the hysteresis compared to the lower branch, and this could be related to the larger oscillation amplitude observed on the upper branch. However, the steep increase of $Q_{\rm eff}$ near the critical point cannot be explained by an increase of the oscillation amplitude, because $A$ stays constant in this region.

Another possible comparison is with parametric amplification and noise squeezing \cite{rugar91p1}. In noise squeezing, by parametric amplification the noise is transferred from an oscillation with a given phase to the one phase-shifted by $90^{\circ}$, the other quadrature. In this way the noise in one quadrature is reduced and the force sensitivity in this quadrature is indeed enhanced while the noise in the other quadrature is increased. Here, in contrast, oscillations of both phases are amplified similarly. The thermal noise in the quadrature that does not self-oscillate is also enhanced, leading to an overall additional thermal noise in this quadrature \cite{poot12p1}. It has been stressed that the overall noise limit of the measurement resulting from both quadratures, is not affected by the enhancement of the $Q$-factor \cite{kobayashi11p1}.

Ultra-low thermal noise measurements enabled by large effective $Q$-factors using the self-excitation mode have been suggested \cite{tamayo05p1}. For scanning force microscopy in liquids, a reduction of noise has been observed \cite{kobayashi11p1}. This noise reduction had been accompanied by a change in the force constant through optomechanical effects. Taking this change in force constant into account, the final result was that the noise remained comparable to the situation without optomechanical resonance. For UHV measurements in contrast, the change in force constant has been found to be negligible \cite{troeger10p1}. Measurements exploiting optomechanical resonances could be useful for scanning force microscopy measurements in special situations: For example, large effective $Q$-factors could be useful for ultra-sensitive measurements of energy dissipation \cite{hoffmann07p1}. Additionally, noise squeezing could be useful for pump-probe-like measurements \cite{schumacher17p1}.

\begin{acknowledgments}
It is a pleasure to thank Denny Koehler for support in the lab and useful discussions. We thank Carsten Henkel for discussions on nonlinear photonic effects. P.~M. and L.~M.~E. kindly acknowledge financial support by the German Science Foundation (DFG) under grant number EN 434/38-1, EN 434/40-1 and MI 2004/3-1. J.~R.-M. and D.~H. thank the DFG for financial support through the Coordinated Research Center (CRC) 1173 ``Wave Phenomena''. R.~H.-V. acknowledges financial support from the European Research Council through the Starting Grant NANOCONTACTS (No. 239838), from the Ministry of Science and Arts, Baden-W\"urttemberg, in the framework of its Schlieben-Lange program and from the DFG through a Heisenberg fellowship.
\end{acknowledgments}

\section*{\label{sec:level6}Data availability statement}

The data that support the findings of this study are available from the corresponding author upon reasonable request.

\nocite{*}
\bibliography{milde}

\end{document}